\documentclass[a4paper,11pt]{article}
\usepackage{pos}

\usepackage{xspace}
\usepackage{mathrsfs}
\usepackage{amsfonts}
\usepackage{array}
\usepackage{epsfig}
\usepackage{rotating}
\usepackage{multirow}

\usepackage{amsmath}    % need for subequations
\usepackage{verbatim}   % useful for program listings
\usepackage{color}      % use if color is used in text
\usepackage{colordvi}
\usepackage{bm}

\usepackage{subfigure}  % use for side-by-side figures
\usepackage{url}
\usepackage{pstricks,rotating, graphicx}
\usepackage{cancel}

\usepackage{tablefootnote}  % needed for tables

\title{NNLO fits of top-quark mass using total, single-differential and double-differential $t\bar{t}+X$ cross-section data
}
\ShortTitle{NNLO $m_t^{\rm{pole}}$ fits  using total, single- and double-differential $t\bar{t}+X$ cross-section data
}

\author[a]{S.~Alekhin}
\author*[a,b,c]{M.V.~Garzelli}
\author[d]{J.~Mazzitelli}
\author[a]{S.-O.~Moch}
\author[a]{O.~Zenaiev}

\affiliation[a]{II. Institut f\"ur Theoretische Physik, Universit\"at Hamburg \\
	Luruper Chaussee 149, D--22761 Hamburg, Germany}
\affiliation[b]{CERN, Department of Theoretical Physics, 
CH--1211 Geneva 23, Switzerland}
\affiliation[c]{Universit\`a degli Studi di Cagliari, Dipartimento di Fisica \\
	Cittadella Universitaria, I--09042 Monserrato, Italy}
\affiliation[d] {Paul Scherrer Institut, CH--5352 Villigen, Switzerland}

\emailAdd{sergey.alekhin@desy.de}
\emailAdd{maria.vittoria.garzelli@desy.de}
\emailAdd{javier.mazzitelli@psi.ch}
\emailAdd{sven-olaf.moch@desy.de}
\emailAdd{oleksandr.zenaiev@desy.de}

\abstract{
We describe the fits of the top-quark mass value at NNLO using as input the double-differential distributions in rapidity and invariant mass of $t\bar{t}$ pairs obtained by the ATLAS and CMS collaborations from unfolding of their experimental data to the parton level, compared to NNLO theory predictions.
We consider different state-of-the-art PDF sets, finding results of the fits compatible among each other within uncertainties.
On the other hand, we observe some tension among the fits to different datasets.
}

\note[]{Report number: CERN-TH-2024-220}

\FullConference{42nd International Conference on High Energy Physics (ICHEP2024)\\
	18-24 July 2024\\
	Prague, Czech Republic\\}

\begin{document}
	\maketitle

\begin{figure}[h!]
\begin{center}
 \includegraphics[width=0.40\textwidth]{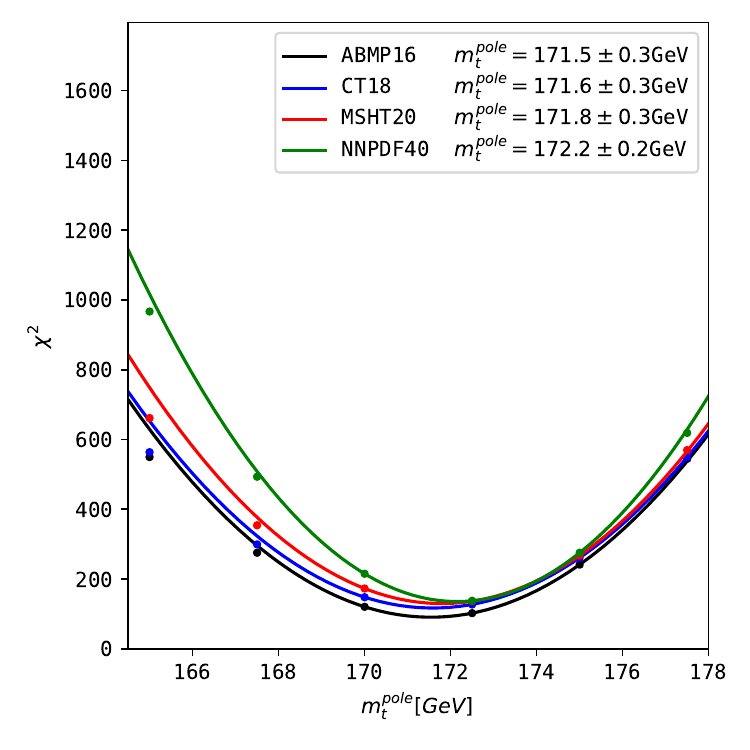}
 \includegraphics[width=0.40\textwidth]{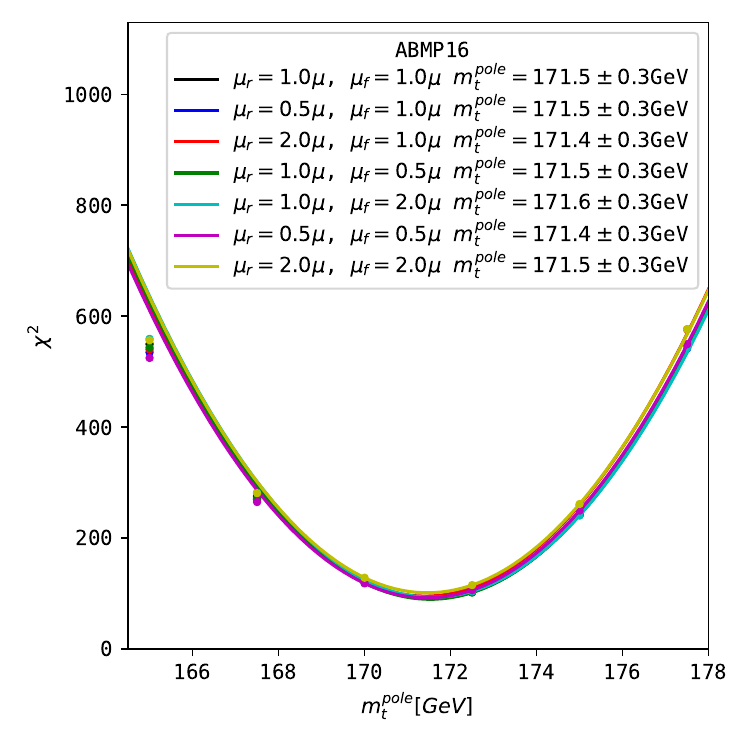}
\caption{\label{fig:chi2} $\chi^2$ for different $m_t^{\rm{pole}}$ values, as well as the best-fit parabolic interpolation to the $\chi^2$ values for $m_t^{\rm{pole}}$~=~170,~172.5,~175~GeV, for different (PDFs+$\alpha_s(M_Z)$) sets (left panel). $\chi^2$ for different ($\mu_r$, $\mu_f$) combinations in the case of the ABMP16 set (right panel).
} 
\end{center}
\end{figure}

\begin{figure}[h]
\begin{center}
 \includegraphics[width=0.325\textwidth]{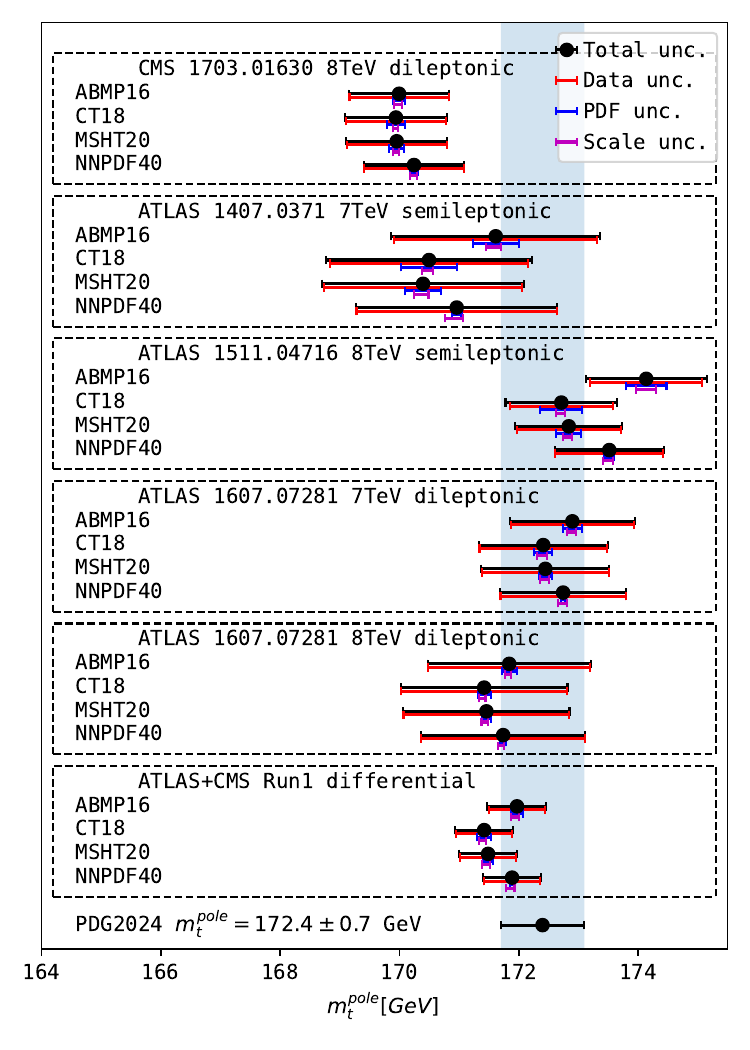}
 \includegraphics[width=0.325\textwidth]{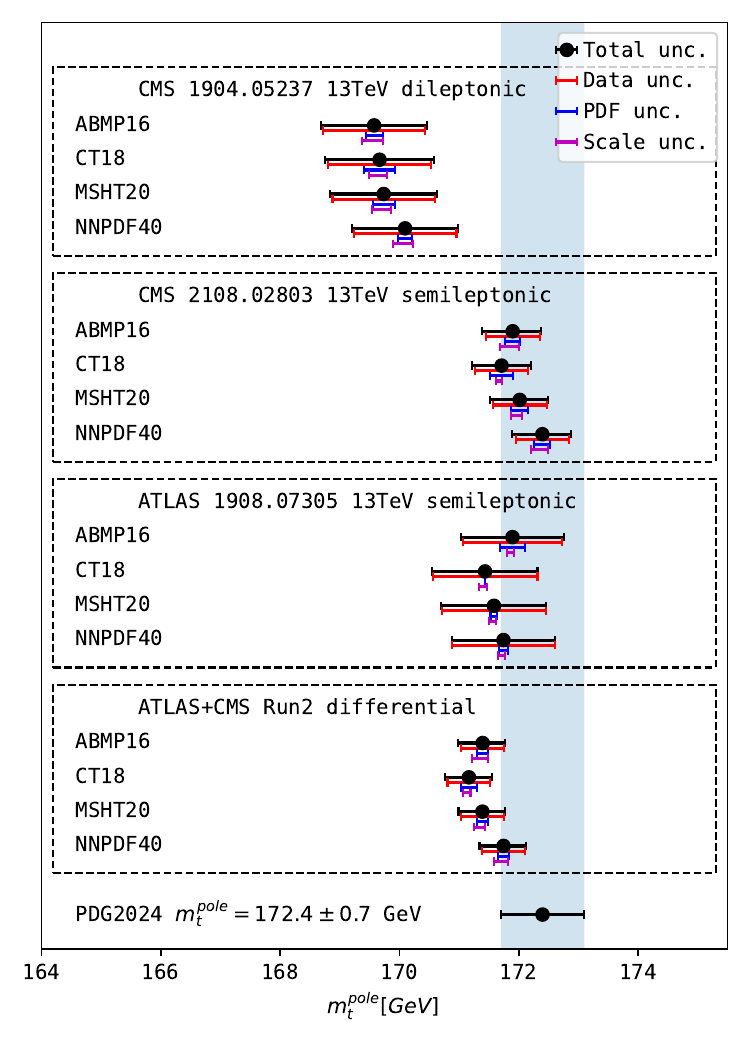}
 \includegraphics[width=0.325\textwidth]{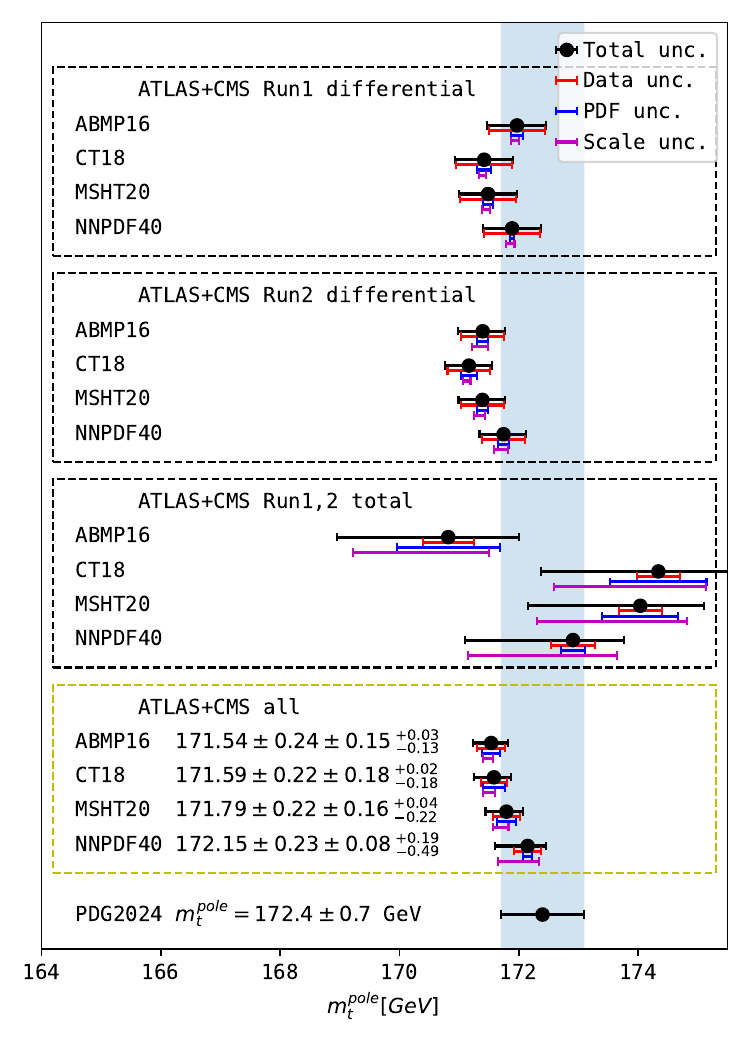}\\
\caption{\label{fig:plotsfit} $m_t^{\rm{pole}}$ values extracted using datasets on normalized (multi-)differential cross sections obtained at the LHC in Run I (left panel) and Run 2 (central panel), as well as datasets from a global analysis of Run 1 + Run 2 differential and total inclusive cross sections (right panel), compared to the PDG 2024 $m_t^{\rm{pole}}$ value. 
%In the latter, 
Our best-fit $m_t^{\rm{pole}}$ values are reported together with uncertainties for each considered PDF~+~$\alpha_s(M_Z)$ set.
}
\end{center}
\end{figure}

The top-quark mass is 
a parameter of the Standard Model. 
Knowing its precise value is relevant for considerations on the stability of our Universe and for the understanding of the Electroweak Symmetry Breaking mechanism. This value has been measured with both direct and indirect techniques 
(for a review see, e.g., Ref.~\cite{Schwienhorst:2022yqu}).
%of these techniques). 
It is possible to extract 
%the top-quark mass value 
it in a well-defined mass renormalization scheme, by
comparing data on $t\bar{t}+X$ hadroproduction cross sections 
to theoretical predictions. 
%In the following we describe 
Our NNLO extraction of the on-shell mass $m_t^{\rm{pole}}$
uses 
as 
ingredients single- and double-differential cross-section measurements by the ATLAS and CMS collaborations, as well as NNLO theoretical predictions computed with an in-house optimized version of the \texttt{MATRIX}~\cite{Catani:2019hip,Grazzini:2017mhc} code, specifically tailored for this process, interfaced to \texttt{PineAPPL}~\cite{Carrazza:2020gss} to facilitate fitting operations. 
We consider as input various state-of-the-art (PDFs+$\alpha_s(M_Z)$) sets. 
More detail is provided in Ref.~\cite{Garzelli:2023rvx} and references therein.
We perform a $\chi^2$ analysis, including in the covariance matrix 
experimental statistical uncertainties, experimental systematic correlated and uncorrelated uncertainties, theoretical uncertainties related to the specific methodology used for computing the cross sections, as well as PDF uncertainties. 
%Unfortunately, the information on correlations of systematic uncertainties of different analyses providing data at the differential level is not available. 
To reduce the effects of the lack of information on correlations of systematic uncertainties of different analyses providing data at the differential level, 
%presently unavailable, 
we use datasets of normalized differential cross sections.  
Scale uncertainties are evaluated separately, building a separate $\chi^2$ for each of the renormalization and factorization scale combinations ($\mu_r, \mu_f$), considering seven-point 
%scale 
variation around a central scale $H_T/4$, and accounting for the spread in the fitted 
%top-quark mass 
$m_t^{\mathrm{pole}}$
values. 
The $\chi^2$'s close to their minima show a parabolic shape for all considered (PDFs+$\alpha_s(M_Z)$) sets and scale combinations. We assume that the top-quark mass value is the minimum of each parabola fitting the $\chi^2$ as a function of $m_t^{\rm{pole}}$ in the three points $m_t^{\rm{pole}} = 170,\, 172.5,\, 175$~GeV. The $\chi^2$'s for the various sets are shown in Fig.~\ref{fig:chi2}, left, together with the best-fit $m_t^{\rm{pole}}$ values. The minima for different sets turn out to be similar. We interpret this as an indication of the robustness of our analysis. In the case of ABMP16~\cite{Alekhin:2017kpj}, the $\chi^2$'s for different ($\mu_r,\mu_f$) combinations are shown in Fig.~\ref{fig:chi2}, right. The uncertainty value $\Delta m_t^{\rm{pole}} = 0.3$~GeV, quoted in the inset of the latter plot, takes into account all uncertainties included in the covariance matrix.
Our main results using different differential datasets in Run~1 and~2, as well as differential and total cross-section datasets from both runs simulteneously, are summarized in the three panels of Fig.~\ref{fig:plotsfit}, respectively.  This shows that Run~2 differential datasets have larger constraining power than Run~1 ones, and the most stringent constraints are played by the dataset of the CMS semileptonic analysis at $\sqrt{S}=13$~TeV~\cite{CMS:2021vhb}, 
based on
the full Run~2 integrated luminosity. 
Datasets for $t\bar{t}$ pairs decaying semileptonically point in general towards $m_t^{\rm{pole}}$ values slightly larger than the dileptonic datasets, but still compatible within 2$\sigma$'s. This tension decreases when performing simultaneous fits of PDFs and top-quark mass values considering their correlations (see e.g.~our recent work Ref.~\cite{Alekhin:2024bhs}). 
On the other hand, the total inclusive cross-section data play a much less relevant role in the fit.
The 
%top-quark pole mass 
$m_t^{\rm{pole}}$ values extracted for all (PDFs+$\alpha_s(M_Z)$) sets turn out to be compatible among each other and also with the PDG 2024 $m_t^{\rm{pole}}$ value~\cite{ParticleDataGroup:2024cfk}. Our best fit values, including the uncertainties, are also reported 
for each set for our most global fit
in the inset in the right panel of Fig.~\ref{fig:plotsfit}.
For the time being, in the case of our most global analysis, data uncertainties amount to $\sim 0.2 - 0.3$~GeV, PDF uncertainties 
%amount 
to $\sim 0.1 - 0.2$~GeV, and NNLO scale uncertainties have 
size similar  to the PDF ones. 
In the near future, data uncertainty 
reduction 
will push for 
theoretical computations beyond NNLO. On the one hand, 
%next-to-next-to-leading logarithmic 
NNLL corrections due to soft-gluon emissions close to threshold can be incorporated in our analysis. On the other hand, approximate N$^3$LO predictions have 
%already 
been obtained for total cross sections and selected differential distributions considering soft-gluon resummation close to threshold~\cite{Kidonakis:2023juy}, but further work is needed to generalize to other distributions, like e.g. those double-differential in $M(t\bar{t})$ and $y(t\bar{t})$ considered in this work, providing sensitivity to the top-quark mass.
\vspace{0.1cm}
\vspace{-0.6cm}
\bibliographystyle{JHEP} 
\bibliography{ttab_ic_new2}
\end{document}